\documentstyle[12pt]{article}
\begin{document}
\vspace*{.3cm}
\begin{center}
{\LARGE\bf{Next-Order Estimate of Higgs $\rightarrow$ Two-Gluon Rate}}\vspace{1cm}\\
{\Large F. A. Chishtie$^*$, V. Elias$^*$, T. G. Steele$^{\dagger}$}\vspace{1cm}\\
{\footnotesize\it $^*$Department of Applied Mathematics, The University of Western Ontario\\
London, Ontario  N6A 5B7  Canada.\\
$^{\dagger}$Department of Physics and Engineering Physics, University of Saskatchewan\\
Saskatoon, Saskatchewan  S7N 5E2  Canada.}
\end{center}
\vspace{-.9cm}
\bigskip
\bigskip
\noindent
{\footnotesize {\bf Abstract.} Using asymptotic Pad\'e approximant methods, we have calculated the
${\cal{O}}(\alpha_s^5)$ contribution to the Standard-Model $H\rightarrow gg$ decay
rate.  This process is of particular interest because of the slow
convergence evident from the known terms of its QCD series.  The ${\cal{O}}(\alpha_s^5)$
term is expressed as a 3rd degree polynomial in $L\equiv ln \left(
\frac{\mu^2}{m_t^2(\mu)}\right)$.  We find that the asymptotic Pad\'e
predictions for the renormalization-group accessible coefficients of
$L$, $L^2$ and $L^3$ are within 1\%, 2\% and 7\% of their respective values.
The inclusion of the ${\cal{O}}(\alpha_s^5)$ contribution
renders the full decay rate virtually scale independent over the entire
$0.3 M_H \leq \mu \leq M_t$ range of the renormalization scale $\mu$.}
\begin{center}
{\large\bf INTRODUCTION}
\end{center}

\indent  The Higgs boson remains the only missing particle within
the Standard
Model of particle physics.  
One of the principal hadronic decay modes for a Weinberg-Salam Higgs (of
mass $100 - 175\, GeV$) is the Higgs $\rightarrow 2$ gluon ($H \rightarrow gg$)
process, a
process whose leading (one-loop) contribution is ${\cal O}(\alpha_s^2)$. The
two known subleading contributions exhibit very slow convergence:
e.g. the decay of an $M_H = 100\,GeV$ Standard-Model Higgs is characterized by
the following
perturbative series[1]:
\begin{eqnarray*}
1 + 0.66 + 0.21.
\end{eqnarray*}
The present paper is directed towards an estimate of the next order (i.e., 
four-loop)
contribution of the $H\rightarrow gg$ process by using  renormalization
group (RG) and asymptotic Pad\'e-approximant methods similar to 
those employed in [2]  for next-order terms in correlation functions,
and in [3] for
the case of a (non-Standard-Model) CP-odd Higgs decay into two gluons.
In the analysis of this latter process, we obtained
Pad\'e-approximant estimates of RG-accessible coefficients of
next-order perturbation theory that are accurate to within a few percent, 
providing support for the 
Pad\'e-predicted RG-inaccessible coefficient still needed for an estimate of the 
${\cal{O}}(\alpha_s^5)$ contribution. In the present work,
we derive the RG accessible coefficients of the Standard-Model $H
\rightarrow gg$ rate within an $\overline{MS}$ framework with
six active flavours [1].  RG invariance allows us to calculate all the coefficients
$c_k$ of $ln^k \left[ \frac{\mu^2}{m_t^2 (\mu)} \right]$ except for the
constant $(k = 0)$ term. We then
demonstrate how Pad\'e approximant methods may be
utilized to estimate the full ${\cal{O}}(\alpha_s^5)$ contribution to
the decay rate, and show that such estimates of the
RG-accessible coefficients are quite close to their
true values. Finally, we discuss how incorporation of the
estimated four-loop term eliminates virtually all residual scale- ($\mu$-)
dependence in the overall $H \rightarrow gg$ rate\vspace{.20cm}.\\
\begin{center}
{\large\bf RG-ACCESSIBLE COEFFICIENTS OF THE $H \rightarrow gg$ RATE}
\end{center}

The $H \rightarrow gg$ decay rate is given by the
following perturbative expression in the $4M_t^2 >> M_H^2 >> M_b^2$ limit [1,4]:
\begin{equation}
\Gamma_{H\rightarrow gg} = \frac{\sqrt{2} G_F M_H^3}{72\pi} S \left[
x^{(6)}(\mu), L(\mu), T \right],
\end{equation}
$$S\left[x,L,T\right]=x^2 \biggl( 
1 + x \left[ \left( \frac{215}{12} -\frac{23T}{6} \right) + \frac{7}{2} L \right] 
\biggr.\nonumber$$
$$+  x^2 \left[ \left( 146.8912 - \frac{4903}{48} T + \frac{529}{48} T^2 \right)\right. \nonumber$$
$$\left. + \left( \frac{1445}{16} - \frac{161}{8} T \right) L + \frac{147}{16} L^2 \right] \nonumber$$
\begin{equation}
\biggl. +  x^3 \left[ c_0 + c_1 L + c_2 L^2 + c_3 L^3 \right] + {\cal{O}}(x^4)\biggr).
\end{equation}

The rate (1) is expressed entirely in terms of a scale-dependent $t$-quark mass
and coupling constant characterised by six active flavours via $L(\mu) \equiv ln \left( \frac{\mu^2}{m_t^2 (\mu)}\right)$ 
and $x^{(6)}(\mu) = \alpha_s^{(6)}(\mu)/\pi$.
The constant $T \equiv ln \left(
\frac{M_H^2}{M_t^2}\right)$ involves a scale-independent ratio of propagator-pole
Higgs and $t$-quark masses [4].
The coefficients $c_i$ in (2) characterize
the unknown 4th loop contribution to the decay
rate.  Three of these unknown coefficients [$c_1, c_2, c_3$] can
be extracted via the RG-invariance of the physical decay rate:
\begin{eqnarray}
0 = \mu^2 \frac{dS}{d\mu^2} [x,L,T], \nonumber \\
= [1 - 2\gamma_m (x)] \frac{\partial S}{\partial L} + \beta(x)
\frac{\partial S}{\partial x} .
\end{eqnarray}
The $\beta$ and $\gamma_m$ functions for six active flavours are given by
\begin{equation}
\beta^{(6)}(x) = -\frac{7}{4} x^2 - \frac{13}{8} x^3 + \frac{65}{128}x^4 ...
\end{equation}
\begin{equation}
\gamma_m^{(6)} (x) = -x - \frac{27}{8} x^2 ...\; \; \;  .
\end{equation}
One can verify that (3) is valid order by
order to known terms in (2).  The continued validity to orders $x^5 L^2$, $x^5
L$ and $x^5$ yield values for $c_1, c_2$ and $c_3$:
\begin{equation}
c_1 = 910.3167 - \frac{16643}{24} T + \frac{3703}{48} T^2,\;
c_2 = \frac{1225}{4} - \frac{1127}{16} T,\;
c_3 = \frac{343}{16}.
\end{equation}
In the following section, we will utilize asymptotic Pad\'e approximation
procedure to estimate $\{c_0, c_1, c_2, c_3 \}$.  The ability to predict
known coefficients will serve to test the appropriateness of this procedure for
estimating the RG-inaccessible coefficient $c_0$\vspace{.20cm}.\\
\begin{center}
{\large\bf PAD\'E PREDICTIONS FOR FOUR-LOOP COEFFICIENTS}
\end{center}
The series $S[x,L,T]$ in (2) may be expressed as follows:
\begin{equation}
S[x,L,T] = x^2 \left[ 1 + R_1 [L,T] x + R_2 [L,T]x^2 + R_3[L,T]x^3 + ... \right],
\end{equation}
where
\begin{equation}
R_1[L,T] = \left( \frac{215}{12} - \frac{23}{6} T \right) + \frac{7}{2}L,
\end{equation}
\begin{equation}
R_2 [L,T] = 146.8912 - \frac{4903}{48}T + \frac{529}{48}T^2 
+ \left( \frac{1445}{16} - \frac{161}{8} T \right)L + \frac{147}{16} L^2,
\end{equation}
\begin{equation}
R_3 [L,T] = c_0(T) + c_1(T)L + c_2(T)L^2 + c_3 L^3 .
\end{equation}

As discussed in prior work [2, 3, 5] and in another contribution
to these Proceedings,\footnote[1]{See ``Reducing the Theoretical Uncertainty
in the Extraction of $|V_{ub}| \ldots$'' in this volume.}
Pad\'{e}-approximant methods may be utilized in order to predict that
\begin{equation}
R_3[L] = 2R_2^3 [L] / (R_1[L] R_2[L] + R_1^3 [L])
\end{equation}
To obtain $c_1, c_2, c_3$, we match the scale dependence of (10) to
that of (11) over the $L > 0$ region [ultraviolet scales of $\mu > M_t
(\mu)]$ by optimizing the least squares function
\begin{equation}
\chi^2 [c_0, c_1, c_2, c_3] = \int_0^1 \left[ R_3 - (c_0 - c_1 ln w +
c_2 ln^2 w - c_3 ln^3 w)\right]^2 dw,
\end{equation}
as in our prior contribution to this Proceedings,$^1$ with
$w = \frac{m_t^2 (\mu)}{\mu^2}$ [$L = -ln (w)]$, and with $R_3$ given
by (11). 
For $M_H = 100\,GeV$ $(T = -1.126)$, we find that
$$\chi^2(c_0,c_1,c_2,c_3) \nonumber$$
$$ =  4.064168878 \cdot 10^7 + 720 c_3^2 
+ 4 c_0 c_2 + c_0^2 + 2c_0 c_1 + 24 c_2^2 
+ 12 c_1 c_2 + 12 c_0 c_3 
+2c_1^2 \nonumber$$
\begin{equation}+  240 c_2 c_3 + 48 c_1 c_3 - 
10205.87 c_0 -17507.05 c_1 - 54106.23 c_2 -
234536.67 c_3 .
\end{equation}
Optimization requires that
\begin{equation}
\frac{\partial \chi^2}{\partial c_i} = 0
\end{equation}
which yields the following predicted values:
\begin{equation}
c_0^{\chi^2} = 2452, \; \; c_1^{\chi^2} = 1774, \; \; c_2^{\chi^2} 
= 377.2, \; \; c_3^{\chi^2} = 20.45,
\end{equation}
The corresponding true values of
the RG-accessible coefficients (6) when
$M_H = 100\,GeV$ are
\begin{equation}
c_1 = 1789.02, \; c_2 = 385.568, \;  c_3 = 21.4375.
\end{equation}
indicative of the accuracy of the estimation procedure leading to (15).
In Table 1, a tabulated set of predictions of the four loop term $c_0$
is given for $M_H$ between 100 and $175\,GeV$. Relative errors of $c_1$
and $c_2$ are seen to remain within 2\% over the range of $M_H$ given;  $c_3$
is seen to be within 7\% of its true value for the same range.  

The results for $c_0$, as tabulated in Table 1, can be utilized to predict $c_0$
as a degree-3 polynomial in $T$:
\begin{equation}
c_0(T) = 755.9 - 1029T + 394.3T^2 - 26.74T^3.
\end{equation}
This prediction, as obtained via Pad\'{e}-approximant
methods, can be coupled with $c_3$ and the known $T$-dependence (6) of $c_1(T)$
and
$c_2(T)$ to predict the full four-loop contribution (10) to
$S[x(\mu), L(\mu), T]$, the scale-sensitive (i.e., $\mu$-dependent)
portion (7) of the $H \rightarrow gg$ rate (1).

Surprisingly, the scale-dependence of $S[x(\mu), L(\mu), T]$ is
virtually eliminated upon incorporation of the estimated four-loop
contribution to the rate. In Table 2, this scale-sensitive portion of
the $H \rightarrow gg$ rate is evaluated for $M_H = 125\,GeV$ for selected values
of the scale parameter $\mu$ between 38 and $175\,GeV$. The four-loop term
$R_3 x^3$, as defined by (10) to include the $M_H = 125\,GeV$ Pad\'{e}
estimate $c_0 = 1646$ from Table 1, varies from zero to 7\% of the
leading term in the perturbative series within $S[x, L, T]$, depending on
the choice for $\mu$. Nevertheless, $S[x, L, T]$ displays a relative
error of 0.6\% over this same range of $\mu$, suggesting that the
Pad\'{e} estimate for $c_0$ is precisely what is required to remove
virtually all residual scale- ($\mu$ -) dependence from the 
four-loop-order decay rate.
\eject
\begin{center}
\begin{tabular}{c|c|c|c|c|c|c|c} 
{\footnotesize 
$M_H(GeV)^1$} & {\footnotesize $c_0^{pr}$}
 & {\footnotesize $c_1^{pr}$} & {\footnotesize $c_1^{RG}$} & 
{\footnotesize $c_2^{pr}$} & 
{\footnotesize $c_2^{RG}$} & {\footnotesize $c_3^{pr}$} & {\footnotesize $c_3^{RG}$} \\
\hline\hline
{\footnotesize 100} & {\footnotesize 2453} & {\footnotesize 1772}
& {\footnotesize 1789} & {\footnotesize 378.4} & {\footnotesize 385.6}
 & {\footnotesize 20.27} & {\footnotesize 21.44}\\
{\footnotesize 125} & {\footnotesize 1646} & {\footnotesize 1405}
 & {\footnotesize 1417} & {\footnotesize 349.6} & {\footnotesize 354.1} & 
{\footnotesize 20.17} & {\footnotesize 21.44}\\
{\footnotesize 150} & {\footnotesize 1120} & {\footnotesize 1126}
 & {\footnotesize 1137} & {\footnotesize 326.2} & {\footnotesize 328.4} 
& {\footnotesize 20.07} & {\footnotesize 21.44}\\
{\footnotesize 175} & {\footnotesize 763.2} & {\footnotesize 903.7} & {\footnotesize 915.1}
 & {\footnotesize 306.6} & {\footnotesize 306.7} & {\footnotesize 19.98} & {\footnotesize 21.44}
\end{tabular}
\end{center}
{\footnotesize
{\bf TABLE 1:} 
Predicted (pr) and true (RG) values (as determined by renormalization-
group methods) for four-loop coefficients of the $H \rightarrow gg$
decay rate, as obtained via the moment approach of ref. [4]. Least-
squares predictions, as in (15) for $M_H = 100\,GeV$ are virtually
identical to those of the moment approach, which also depend on
the asymptotic Pad\'{e} prediction (11) for the four-loop term\vspace{.8cm}.}\\
\begin{center}
\begin{tabular}{c|c|c|c|c}
{\footnotesize $\mu$} & {\footnotesize $\alpha_s(\mu)$} 
& {\footnotesize $m_t(\mu)$} & {\footnotesize $1 + R_1x + R_2x^2 + R_3x^3$} &
{\footnotesize $S[x(\mu), L(\mu), T]$}\\
{\footnotesize $(GeV)$} &                 &{\footnotesize $(GeV)$}       &                               & \\
\hline\hline
{\footnotesize 38} & {\footnotesize 0.1353} & {\footnotesize 200.9} 
& {\footnotesize 1 + 0.3818 - 0.0439 + 0.0026} & {\footnotesize 0.002486}\\
{\footnotesize 66} & {\footnotesize 0.1245} & {\footnotesize 190.6}
 & {\footnotesize 1 + 0.5192 + 0.0658 - 0.0000} & {\footnotesize 0.002490}\\
{\footnotesize 80.5} & {\footnotesize 0.1211} & {\footnotesize 187.3} 
& {\footnotesize 1 + 0.5631 + 0.1067 + 0.0074} & {\footnotesize 0.002491}\\
{\footnotesize 125} & {\footnotesize 0.1141} & {\footnotesize 180.4} 
& {\footnotesize 1 + 0.6526 + 0.1986 + 0.0367} & {\footnotesize 0.002486}\\
{\footnotesize 175} & {\footnotesize 0.1092} 
& {\footnotesize 175.6} & {\footnotesize 1+ 0.7127 + 0.2668 + 0.0676} 
& {\footnotesize 0.002475}
\end{tabular}
\end{center}
{\footnotesize {\bf TABLE 2:} 
Virtual scale invariance of $H \rightarrow gg$ four-loop estimated rate
when $M_H = 125\,GeV$. $\alpha_s(\mu)$ is assumed to evolve from
$\alpha_s(M_Z) = 0.119$ [6], and $m_t(\mu)$ is assumed to evolve from
$m_t(m_t) \cong M_t = 175.6\,GeV$. The order-by-order perturbation series
within the decay rate, as defined by (7), as well as $S[x(\mu), L(\mu),
T]$, the scale-sensitive portion of the rate (1), are listed for
representative values of $\mu$ between $M_H/3$ and $M_t$. Although the
predicted four loop term $R_3x^3$ varies between zero and seven percent
of the lead (one-loop) term in the series, the scale-sensitive portion
of the rate $S[x, L. T]$ is constant up to $|\Delta S/S| = 0.6\%$\vspace{.8cm}.}
\begin{center}
\begin{tabular}{c|c|c|c|c}
{\footnotesize $\mu$} & {\footnotesize $\alpha_s(\mu)$} & {\footnotesize $m_t(\mu)$} 
& {\footnotesize $1 + R_1x + R_2x^2 + R_3x^3$} &
{\footnotesize $S[x(\mu), L(\mu), T]$}\\
{\footnotesize $(GeV)$} &                 &{\footnotesize $(GeV)$}       &                               & \\
\hline\hline
{\footnotesize 47} & {\footnotesize 0.1309} & {\footnotesize 196.8} 
& {\footnotesize 1 + 0.3793 - 0.0378 + 0.0023} & {\footnotesize 0.002334}\\
{\footnotesize 78.5} & {\footnotesize 0.1215} & {\footnotesize 187.7} 
& {\footnotesize 1 + 0.5036 + 0.0589 + 0.0000} & {\footnotesize 0.002337}\\
{\footnotesize 95.5} & {\footnotesize 0.1183} & {\footnotesize 184.6} 
& {\footnotesize 1 + 0.5463 + 0.0973 + 0.0064} & {\footnotesize 0.002338}\\
{\footnotesize 150} & {\footnotesize 0.1114} & {\footnotesize 177.8} 
& {\footnotesize 1 + 0.6360 + 0.1866 + 0.0333} & {\footnotesize 0.002334}\\
{\footnotesize 175} & {\footnotesize 0.1092} & {\footnotesize 175.6} 
& {\footnotesize 1+ 0.6641 + 0.2170 + 0.0458} & {\footnotesize 0.002330}
\end{tabular}
\end{center}
{\footnotesize {\bf TABLE 3:} 
Virtual scale invariance of $H \rightarrow gg$ four-loop estimated rate
when $M_H = 150\,GeV$. Running of $\alpha_s(\mu)$ and $m_t(\mu)$, as well
as the perturbative series $1 + R_1 x + \ldots$ within $S[x(\mu),
L(\mu), T]$, the scale-sensitive portion of the rate (1), are as in Table
2. Although the four-loop term $R_3x^3$ grows to be 4.6\% of the leading
term (unity) in the series, the rate itself is seen to exhibit a
relative error $|\Delta S/S| \stackrel{<}{_\sim} 0.3\%$ over the range $M_H/3 
\stackrel{<}{_\sim} \mu
\stackrel{<}{_\sim} M_t$\vspace{.8cm}.}
\begin{center}
\begin{tabular}{c|c|c|c|c}
{\footnotesize $M_H$} & {\footnotesize $\mu^{(MS)}$} & {\footnotesize $\delta_m(\mu^{MS})$}
 & {\footnotesize $R_3x^3(\mu^{MS})$} &
{\footnotesize $\Gamma(H \rightarrow gg)$}\\
{\footnotesize $(GeV)$} & {\footnotesize $(GeV)$}        &            &                               
& {\footnotesize $(GeV)$}\\
\hline\hline
{\footnotesize 100} & {\footnotesize 65.5} 
& {\footnotesize -0.059} &  {\footnotesize + 0.0089} 
& {\footnotesize $1.90 \cdot 10^{-4}$}\\
{\footnotesize 125} & {\footnotesize 80.5} & {\footnotesize -0.015}
 & {\footnotesize + 0.0074}  & {\footnotesize $3.52 \cdot 10^{-4}$}\\
{\footnotesize 150} & {\footnotesize 95.5} & {\footnotesize +0.023} 
&  {\footnotesize + 0.0064} & {\footnotesize $5.83 \cdot 10^{-4}$}\\
{\footnotesize 175} & {\footnotesize 111} & {\footnotesize +0.077}
 & {\footnotesize + 0.0309}  & {\footnotesize $9.08 \cdot 10^{-4}$}
\end{tabular}
\end{center}
{\footnotesize {\bf TABLE 4:} 
Incorporation of Mass-corrections $\delta_m(\mu^{MS})$ to the
perturbative series (22) in the predicted Standard-Model $H \rightarrow
gg$ decay rate for various choices of $M_H$, as described in the text.
The ``minimal sensitivity'' choice of scale $(\mu^{MS})$ represents a
(very weak) local maximum of $S[x(\mu), L(\mu), T]$ prior to the
incorporation of mass-corrections. The magnitude of the leading mass
correction $\delta_m$ is shown to be several times larger than the
estimated four-loop correction $R_3x^3$, although it is generally
smaller than the known three-loop correction $R_2x^2$\vspace{.8cm}.}\\
This reduction in scale dependence is even more dramatic as $M_H$
increases to $150\,GeV$, as is evident from Table 3.  Incorporation of the
Pad\'{e} estimate $c_0 = 1120$ (Table 1) into $R_3 x^3$, as defined by
(10) [with RG-values (6) for $\{c_1, c_2, c_3\}$], leads to a four-loop
expression for the rate 
characterised by a relative error $|\Delta S/S|
\leq 0.3\%$ over values of the scale parameter $\mu$ chosen between 47
and $175\,GeV$. It must be noted that $c_0$'s contribution to $S[x(\mu),
L(\mu), T]$ is just $c_0 x^3(\mu)$, in itself an appreciably scale-dependent contribution. 
The absence of any concomitant scale dependence
in $S$, despite its potentially large contribution from an arbitrary
choice of $c_0$, demonstrates the success of the Pad\'{e} estimate for $c_0$ in
controlling potentially large residual scale sensitivity in the $H
\rightarrow gg$ rate\vspace{.20cm}.
\begin{center}
{\large\bf LEADING FERMION MASS CORRECTIONS}
\end{center}

The rate (1) has implicit within its lead term the assumption that
$M_H^2 << 4M_t^2$ [1]. Physical departures from this assumption, while
small compared to the three-loop contribution to the $H \rightarrow gg$
rate as calculated in [1], are in fact somewhat larger than the four-loop 
contributions estimated above. These fermion mass corrections are
most easily incorporated by replacing the leading factor of unity in the
series $1 + R_1x + R_2x^2 + R_3x^3$ with the following departure from
the assumed $m_b^2 << M_H^2 << 4M_t^2$ hierarchy of masses [4,7]:
\begin{equation}
1 \rightarrow \frac{9}{16} \left[ (A_t+Re A_b)^2 + (Im A_b)^2 \right]
\equiv 1 + \delta_m (\mu)
\end{equation}
\begin{equation}
A_t = 2\left[ \tau_t + (\tau_t - 1) \left(sin^{-1}(\sqrt{\tau_t}) 
\right)^2 \right] / \tau_t^2, \; \; \; \tau_t \equiv M_H^2 / 4m_t^2(\mu),
\end{equation}
\begin{equation}
A_b = 2 \left[ \tau_b + (\tau_b - 1) f (\tau_b) \right] / \tau_b^2, \; \; \; \tau_b 
\equiv M_H^2 / 4m_b^2(\mu),
\end{equation}
\begin{equation}
f(\tau) \equiv - \frac{1}{4} \left[ ln \left( \frac{1+\sqrt{1-1/\tau}}
{1- \sqrt{1-1/\tau}} \right) - i \pi\right]^2.
\end{equation}
We (somewhat arbitrarily) choose the scale $\mu$ to be the ``minimal-
sensitivity'' [8] maximum ($\mu^{MS}$) of $S[x(\mu), L(\mu), T]$, which
is seen to occur at values of $\mu$ tabulated in Table 4 for various
choices of $M_H$. The values for $m_b(\mu^{MS})$ and $m_t(\mu^{MS})$
utilized in (19) and (20) are evolved from refernce values $m_t(m_t) =
175.6\,GeV$, $m_b(m_b) = 4.2\,GeV$ [6]. The leading mass correction
$\delta_m(\mu^{MS})$, as obtained from (18-21), is tabulated and compared
to the corresponding estimate of the four-loop term $R_3 x^3(\mu^{MS})$
in Table 4. These mass corrections are clearly seen to dominate over the
four-loop term in the scale-sensitive portion of the $H \rightarrow gg$
rate, which is now given by
$$S[x(\mu^{MS}),L(\mu^{MS}), T]\nonumber $$
\begin{equation}
= x^2(\mu^{MS})\left[1 + \delta_m(\mu^{MS}) + R_1x(\mu^{MS}) +
R_2x^2(\mu^{MS}) + R_3x^3(\mu^{MS})\right].
\end{equation}
Predictions for the full Standard Model $H \rightarrow gg$ rate that are
inclusive of these mass corrections are also listed in Table 4. Although
these leading mass corrections are two to six times as large as the
estimated four-loop correction, they are seen to be smaller (in general)
than the three-loop corrections $R_2x^2(\mu^{MS})$. Moreover, subsequent {\it
subleading} departures from the $m_b^2 << M_H^2 << 4M_t^2$ mass
hierarchy are anticipated to be ${\cal O}(x\delta_m)$, which should
be comparable to or somewhat smaller than the four-loop terms listed in
Table 4. Note that $|\delta_m(\mu^{MS})|$ is itself only $\sim 2\%$ of
the leading term (i.e., unity) in the perturbative series for the $M_H =
125$ and $150\,GeV$ cases, as evident from Table 4, thereby providing
justification for the three-digit accuracy of the $H \rightarrow gg$
predicted rates for these cases\vspace{.15cm}.\\
\begin{center}
{\large\bf REFERENCES}
\end{center}
{\footnotesize{
1.  Chetyrkin, K.G., Kniehl, B.A., and Steinhauser, M.,
{\it Phys. Rev. Lett.} {\bf 79}, 353 (1997).\\
2. Chishtie, F., Elias, V., and Steele, T.G. {\it Phys.
Rev. D} {\bf 59}, 105013 (1999).\\
3. Chishtie, F.A., Elias, V., and Steele, T.G., {\it J.
Phys. G} {\bf 26}, 93 (2000).\\
4. Chishtie, F.A., Elias, V., and Steele, T.G., {\it J. Phys.
G} {\bf 26} (to appear: hep-ph/0004140).\\
5. Elias, V., Steele, T.G., Chishtie, F., Migneron, R., and
Sprague, K., {\it Phys. Rev. D} {\bf 58}, 116007 (1998).\\
6. Caso, C. et. al. [Particle Data Group], {\it Eur. Phys. J.
C} {\bf 3}, 1 (1998).\\
7. Spira, M., Djouadi, A., Graudens, D., and Zerwas, P.M.,
{\it Nucl. Phys. B} {\bf 453}, 17 (1995).\\
8. Stevenson, P.M., {\it Phys. Rev. D} {\bf 23}, 2916 (1981).}}

\end{document}